\documentclass[epj]{svjour}

\makeatletter
\let\cl@chapter\undefined
\makeatletter

\usepackage[utf8]{inputenc}
\usepackage{amsmath}
\usepackage{amsfonts}
\usepackage{amssymb}
\usepackage{graphicx}
\usepackage{color}
\usepackage{hyperref}
\usepackage{xspace}
\usepackage{dcolumn}
\usepackage{cleveref}
\usepackage{grffile}
\usepackage{epstopdf}

\usepackage{microtype}

\newcommand{\HS}{Hagedorn state\xspace}
\newcommand{\HSS}{Hagedorn states\xspace}
\newcommand{\HT}{Hagedorn temperature\xspace}

\newcommand{\dd}{{\rm d}}

\newcommand{\eg}{\textit{e.g.}\xspace}
\newcommand{\ie}{\textit{i.e.}\xspace}

\newcommand{\UNIT}[1]{\ensuremath{\,{\rm #1}}\xspace}

\newcommand{\MeV}{\UNIT{MeV}}
\newcommand{\GeV}{\UNIT{GeV}}

\newcommand{\fm}{\UNIT{fm}}
\newcommand{\mb}{\UNIT{mb}}
\newcommand{\proz}{\UNIT{\%}}

\DeclareMathOperator{\BessK}{K}

\newcommand{\K}[1]{\BessK_{#1}}

\definecolor{magenta}{cmyk}{0,1,0,0}

\newcommand{\REM}[1]{}

\renewcommand{\vec}[1]{\mathbf{#1}}

\newcommand*{\defeq}{\mathrel{\vcenter{\baselineskip0.5ex \lineskiplimit0pt
                     \hbox{\scriptsize.}\hbox{\scriptsize.}}}%
                     =}

\begin{document}

\title{Production of Light Nuclei in Heavy Ion Collisions via Hagedorn Resonances}

\author{K.~Gallmeister \and C.~Greiner}
\institute{Institut f\"ur Theoretische Physik, Goethe-Universit\"at
  Frankfurt am Main, Max-von-Laue-Str.~1, 60438 Frankfurt am Main,
  Germany}

\date{Received: date / Revised version: date}

\abstract{
  The physical processes behind the production of light nuclei in heavy ion collisions are unclear. The nice theoretical description of experimental yields by thermal models conflicts with the very small binding energies of the observed states, being fragile in such a hot and dense environment. Other available ideas are delayed production via coalescence, or a cooling of the system after the chemical freeze-out according a Saha equation, or a `quench' instead of a thermal freeze-out.
  A recently derived prescription of an (interacting) Hagedorn gas is applied to consolidate the above pictures.
  The tabulation of decay rates of \HSS into light nuclei allows to calculate yields usually inaccessible due to very poor Monte Carlo statistics. Decay yields of stable hadrons and light nuclei are calculated.
  While the scale-free decays of \HSS alone are not compatible with the experimental data,
  a thermalized hadron and \HS gas is able to describe the experimental data.
  Applying a cooling of the system according a Saha-equation with conservation of nucleons and anti-nucleons in number leads to (nearly) temperature independent yields, thus a production of the light nuclei at temperatures much lower than the chemical freeze-out temperature is possible.
  \PACS{
      {24.10.Pa}{Thermal and statistical models}   \and
      {25.75.-q}{Relativistic heavy-ion collisions}
     } 
}

\maketitle

\section{Introduction}
\label{sec:Introduction}

In recent years, the production of light nuclei in (ultra-) relativistic heavy ion collisions has gained new interest. Experimental measurements of the production of deuteron, triton, helium-3 and helium-4, their anti-particles, and also hyper-triton in high-energetic collisions by the ALICE collaboration at the LHC or some subset of these nuclei in low-energetic collisions by the HADES collaboration at GSI introduce a fundamental question onto their production mechanism.  It is unclear, why the experimental yields can be described so well by thermal models as \eg shown in \cite{Andronic:2010qu,Andronic:2017pug,Lorenz:2017xfn}.

Under the assumption, that a thermalized system has been built up, the binding energies of the observed states are so small, that a survival in such a virulent system of such fragile states at the chemical freeze-out temperatures of ${\cal O}(150\MeV)$ is improbable. Therefore a later production of these nuclei in the time evolution of the collision may be some explanation.

Here the first ansatz is, that in the framework of coalescence, the production of high-mass resonances is governed by the yields of the lower mass states \cite{Scheibl:1998tk,Mrowczynski:2016xqm,Zhao:2018lyf,Bellini:2020cbj}, while still energy conservation is not given in this picture.

Another explanation relies on the assumption of detailed balance, resp.~the law of mass action, resp.~a kind of Saha-equation, which dictates the yields at later stages already by the chemical-freeze-out conditions of the stable hadrons \cite{Vovchenko:2019aoz}. Adjusting chemical potentials have also been introduced in \cite{Xu:2018jff}.

Recently, the additional idea has been discussed, that all these observed yields do not originate from a thermalized gas after a phase transition, but are generated by a `quench' into a state described by \HSS and their decays \cite{Castorina:2019pnb}. Here the underlying picture is a so-called `self organized criticality' (SOC). Thus, instead being in a thermalized and stable state, the system is assumed to be in a critical state, where modifications in all extensions are possible, but keeping the system in its (critical) state, and it just looks like it would be in a stable state.

In refs.~\cite{Beitel:2014kza,Beitel:2016ghw} the authors have developed a prescription of a microcanonical bootstrap of \HSS with the explicitly conserved baryon number $B$, strangeness $S$ and electric charge $Q$, which has been augmented by the consideration of $B$, $S$ and isospin $I$ in \cite{Gallmeister:2017ths}. It is a reformulation of the original concept by Hagedorn himself \cite{Hagedorn:1965st} according to Frautschi \cite{Frautschi:1971ij}, where the covariant formulation is analogous to \cite{Hamer:1972wz,Yellin:1973nj}.

We are thus in the favorable situation to test the above assumptions against experimental data. We will therefore first show, that the \HSS defined in our prescription indeed (nearly) produce a scale independence concerning their decay branching ratios. Nevertheless, these decays modestly fail to describe the experimental yields. On the other hand the assumption of a thermalized system of hadrons together with \HSS leads to a satisfactory description of the experimental data. Whether it was really a thermal system at the freeze-out temperature, or a much cooler system following a Saha equation, which finally produced the observed particles, is not distinguishable within our framework. Thus the criticism against thermal models by confronting low binding energies with large temperatures is not legitimate in our approach.

The paper is organized as follows. We start to recapitulate the basics of the present \HS prescription and elaborate on the extensions needed for the inclusion of light nuclei. Then we first show the decay multiplicities assuming a fixed mass \HS and second, after decays of a thermal \HS gas.  Finally, we discuss the effect of cooling the \HS gas under the assumption of holding yields constant according to the Saha equation.

\section{Hagedorn description}
\label{sec:Model}

We use the microscopic prescription developed in \cite{Beitel:2014kza,Beitel:2016ghw} in its improved formulation described in \cite{Gallmeister:2017ths}. In order to pursue the extensions needed for the light nuclei, we will here first repeat the basic equations as given in \cite{Gallmeister:2017ths}, which are implemented into the transport framework GiBUU \cite{Buss:2011mx}.

Under the basic assumption, that only subsequent two-particle decay participate, the bootstrap equation to be used is
\begin{equation}
\label{eq:basic1}
  \begin{split}
    \tau_{\vec C}(m)&=\tau^0_{\vec C}(m)+\frac{V(m)}{(2\pi)^2}\,\frac1{2m} \ \sideset{}{^*}\sum_{\vec C_1\vec C_2}\iint\dd m_1\dd m_2\\&\times\,\tau_{\vec C_1}(m_1)\tau_{\vec C_2}(m_2) \,m_1\,m_2\,p_{\rm cm}(m,m_1,m_2)\, ,
  \end{split}
\end{equation}
which describes, how the mass degeneration spectrum of the \HSS $\tau_{\vec C}(m)$ is built up from a low mass input $\tau^0_{\vec C}(m)$ and the combination of two lower lying \HSS.  Here, as usual, $4m^2p_{\rm cm}^2=(m^2-m_1^2-m_2^2)^2-4m_1^2m_2^2$, and the special notation $\sum^*$ indicates, that the sum only runs over `allowed' quantum number combinations; $\tau^0_{\vec C}(m)$ stands for the inhomogeneity, \ie the hadronic input, while the volume $V(m)\equiv V$ is just a constant at the moment.  The quantum number vector $\vec C$ may stand for $(BSQ)$ or $(BSI)$ with $B$,$S$,$Q$,$I$ indicating baryon number, strangeness, electrical charge, isospin. As elaborated in \cite{Gallmeister:2017ths}, the combination $(BSI)$ is fully equivalent to $(BSQ)$, but preferable internally.

Selecting different values for the radius, $R$, and thus, via $V=4\pi/3 R^3$, also for the volume $V$ of the \HSS in the bootstrap equation \cref{eq:basic1} yields different slopes and thus different values of the \HT as an intrinsic parameter; larger radii yield steeper increase of the spectrum, thus smaller values of the \HT. The default value $R=1.0\fm$ corresponds to $T_H=167\MeV$, while $R=1.2\fm$ yields $T_H=152\MeV$ (cf.~also \cite{Beitel:2014kza}).

We extend the prescription by the inclusion of light nuclei as stable particles in the input to the bootstrap.
Details of the particles are listed in \cref{tab:nucleiprop}.
\begin{table}
  \begin{center}
    \begin{tabular}{lccccc}
      \hline
      \textbf{}&mass&$B$&$J$&$I$&$S$ \\
      \textbf{}&[GeV]&&&& \\
      \hline
      d$=^{2}$H$       $  & 1.876 &  2 &  1 &  0 &  0 \\
      t\,$=^3$H$,\quad ^3$He$\quad  $  & 2.809 &  3 &  1/2 &  1/2 &  0 \\
      \ $\quad ^3_\Lambda$H$ $  & 2.992 &  3 &  1/2 &  0 & -1 \\
      \ $\qquad\alpha=^4$He$        $ &  3.728 &  4 &  0 &  0 &  0\\
      \hline
    \end{tabular}

    \caption{
      Properties of light nuclei. Listed are baryon number $B$, spin $J$, isospin $I$, and strangeness $S$.
    }

    \label{tab:nucleiprop}
  \end{center}
\end{table}
The resulting Hagedorn spectrum is only very slightly influenced by this addition and the differences are hardly visible. Nevertheless, decays of high mass \HSS now may end in light nuclei as final particles.

In the spirit of ref.~\cite{Vovchenko:2020dmv}, also the inclusion of non-stable resonances could be in order. For this, one would first include these resonances into the Hagedorn bootstrap as if they would also be stable particles. In a second step one then would extend the transport code to implement their decays into stable nuclei and hadrons, as also the decays of hadronic resonances are treated. At the moment, this implies deeper modifications of the algorithm itself and is left for future studies.

It is favorable for the $(BSI)$ prescription, that all of the light nuclei are realized in their lowest isospin level, \ie $I=0$ or $I=1/2$. The fact, that $^3$H and $^3$He are two different charge states in a $I=1/2$ system has to be respected when multiplicity of a special isospin state is calculated.

Identifying particles only according their isospin value does obviously not allow to respect modifications of the wave function, which may be given by details of the ingredients, as \eg their charge states. Like the assumption of a common volume of all \HS specific details between different particle yields are not accessible within our prescription.

The second basic equation is the connection of the decay width $\Gamma$ of some \HS with its production cross section $\sigma$, which is given by \cite{Gallmeister:2017ths}
\begin{equation}
  \label{eq:basic2}
  \begin{split}
    \Gamma_{\vec C}(m)&=\frac{\sigma(m)}{(2\pi)^2}\,\frac1{\tau_{\vec C}(m)-\tau^0_{\vec C}(m)}
    \ \sideset{}{^*}\sum_{\vec C_1\vec C_2}\iint\dd m_1\dd m_2\\
    &\times\,\tau_{\vec C_1}(m_1)\tau_{\vec C_2}(m_2)
    \,p_{\rm cm}^2(m,m_1,m_2)\ .
  \end{split}
\end{equation}
At the moment, the cross section is assumed to show no mass dependence or some other details and is assumed to be a constant. In the actual prescription, it is also directly connected with the radius of the \HS by $\sigma(m)\equiv\sigma=\pi R^2$, \ie $\sigma=31.4\mb$ for $R=1.0\fm$.

Due to the tiny decay probabilities into light nuclei, Monte Carlo studies of the decay chain are not feasible. Looking only at the multiplicities of these light nuclei in the decays, it is possible to tabulate these values. This is analogous to the calculation of the bootstrap/the decay width itself.

Starting from the expression of the calculation of the total decay width \cref{eq:basic2}, (differential) partial branching ratios may be defined by dividing every summand of this expression by its total,
\begin{equation}
  \begin{split}
    &\dd B_{\vec C;\vec C_1,\vec C_2}(m,m_1,m_2)\\
    &\defeq\frac{\dd m_1\dd m_2\,\tau_{\vec C_1}(m_1)\tau_{\vec C_2}(m_2)\,p_{\rm cm}^2(m,m_1,m_2)}
    {\sideset{}{^*}\sum_{\vec C_1\vec C_2}\iint\dd m_1\dd m_2\,\tau_{\vec C_1}(m_1)\tau_{\vec C_2}(m_2)\,p_{\rm cm}^2(\dots)}\ ,
  \end{split}
\end{equation}
such that $\sideset{}{^*}\sum_{\vec C_1\vec C_2}\iint \dd B \equiv 1$.
It is interesting to observe, that here for the \textit{relative} branching ratios, contrary to the decay width \cref{eq:basic2}, the cross section $\sigma(m)$ completely drops out.
The number of a specific light nucleus (${\rm A}={\rm d,\ t,\dots}$, cf.~\cref{tab:nucleiprop}) a given \HS finally decays into is calculated as
\begin{equation}
  \label{eq:precalc}
  \begin{split}
    n^{\rm (A)}_{\vec C}(m)&=\sideset{}{^*}\sum_{\vec C_1\vec C_2}\iint
    \dd B_{\vec C;\vec C_1,\vec C_2}(m,m_1,m_2)\\&\qquad\qquad\times\,\left(n^{\rm (A)}_{\vec C_1}(m_1)+n^{\rm (A)}_{\vec C_2}(m_2)\right)\ .
  \end{split}
\end{equation}
For this purpose, one has to initialize the input correctly, as \eg 
\begin{equation}
  \begin{split}
    n^{\rm (d)}_{(2,0,0)}(m)&\defeq\delta(m-1.876 \,\GeV)\ ,\\
    n^{\rm (t)}_{(3,0,0.5)}(m)&\defeq\delta(m-2.809\, \GeV)\ ,\\
    &\dots\ .
  \end{split}
\end{equation}
This quantity $n^{\rm (A)}_{\vec C}(m)$ gives the total fraction for the decay into the light nucleus, \ie the direct decay and also the indirect decay chain via intermediate \HSS.

The tabulation has to be done for all quantum numbers and masses of the mother particle.
As long as the 'final state' $A$ has fixed quantum numbers and mass, as \eg the states listed in \cref{tab:nucleiprop}, the tabulation is manageable. When looking for an extension of this tabulation to more states, the major problem will be a mass distribution of the final states. In this case, the tabulation will very soon exceed actual memory setups of the HPC computer clusters. Thus a na\"{i}ve extension of \cref{eq:precalc}, especially in the spirit of ref.~\cite{Vovchenko:2019aoz}, is not possible.

\section{Hagedorn decay cascade}
\label{sec:cascade}

The actual implementation of the Hagedorn bootstrap explicitly respects conservation of the quantum numbers.  It is obvious, that the quantum numbers of the initial (mother) state directly influences the yields of the different (daughter and grandchild) states with different quantum numbers.  (As an example, starting with a \HS with $B=2$ yields obviously and considerable more nucleons than starting with $B=0$.)

\begin{figure}[htb]
  \begin{center}
    \hspace*{\fill}%
    \includegraphics[width=0.95\columnwidth,clip=true]{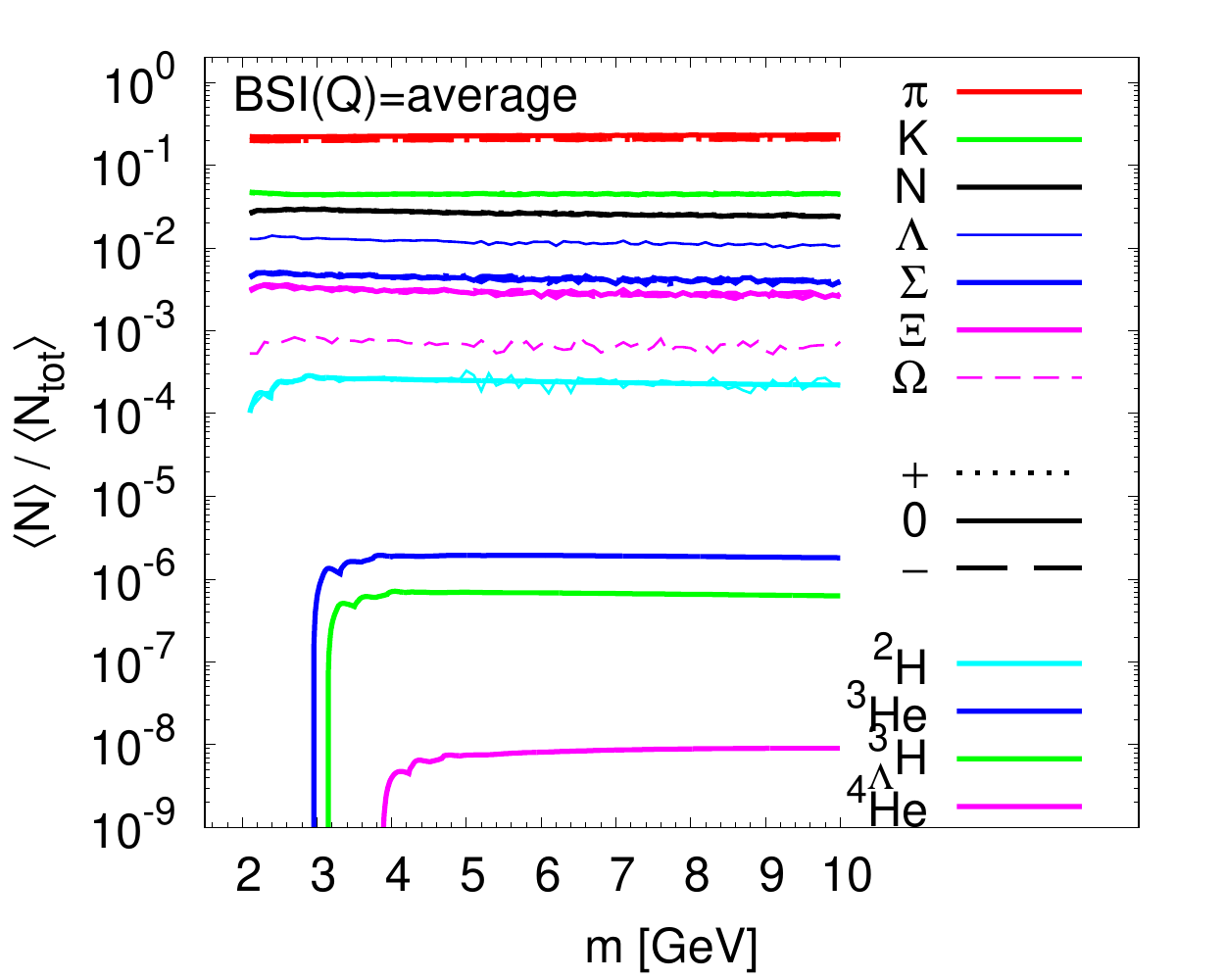}
    \hspace*{\fill}%

    \caption{
      The relative multiplicity of several species as function of the \HS mass $m$ for $T_H=167\MeV$. Different line colors indicate different daughter particles, while line styles as indicated in the plot show different (electrical) charge states.
    }
    \label{fig:Multrel}
  \end{center}
\end{figure}

Calculating `stochastic' averages (contrary to `statistical' averages, which have thermal weights and chemical potentials), one has to average over all possible quantum numbers given by $\tau_{\vec C}$ alone.  While averaging over all quantum numbers which are accessible for a given \HS mass $m$, one observes two general features:
\begin{itemize}
\item The overall yield grows linearly with the \HS mass according $\langle N_{\rm tot}\rangle\simeq 0.27+1.44\GeV^{-1}m$ (see also \cite{Beitel:2014kza})
\item The relative yields are rather independent on the \HS mass, but obey mass thresholds.
\end{itemize}
The latter is illustrated in \cref{fig:Multrel}.
Here all the results for hadronic states are calculated by MC runs, while the yields for the light nuclei are generated by the tabulation described above. For d$=^2$H, results from both approaches are available, match identically, and prove the correctness of the tabulation approach \cref{eq:precalc}. Albeit \cref{fig:Multrel} also shows the different electric-charge states separately, only for pions a slight difference between the charge states is visible. This is due to non-isospin symmetric decay channels of the hadronic resonances.

Of course one has to take the previous statements about the scaling bahavior with some grain of salt, since they rely on figures with logarithmic axis scaling. Nevertheless, for large masses it seems hard to deduce the mass of the mother particle just from \textit{relative} yields.

\begin{figure}[htb]
  \begin{center}
    \hspace*{\fill}%
    \includegraphics[width=0.95\columnwidth,clip=true]{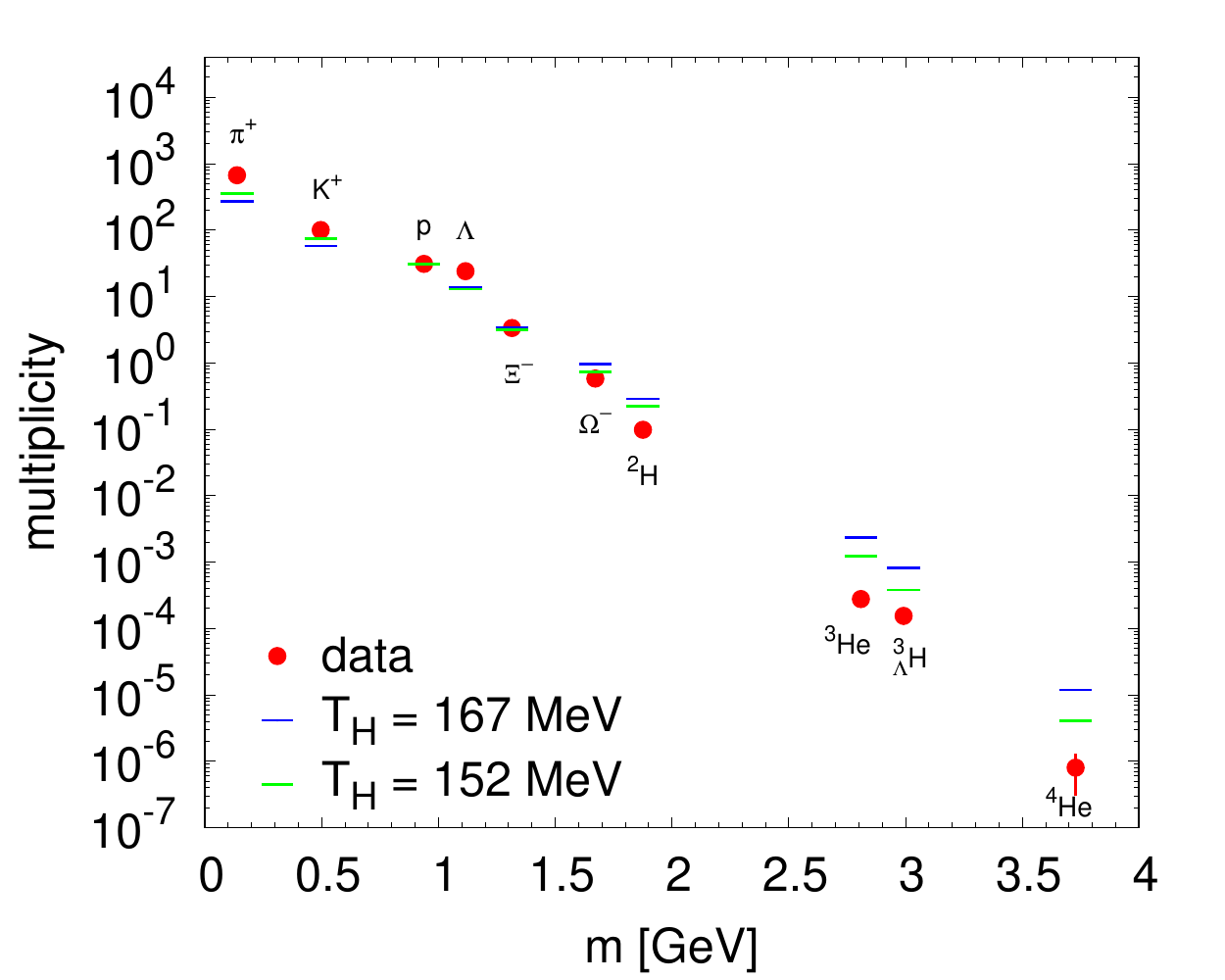}
    \hspace*{\fill}%

    \caption{
      The multiplicities of given stable particles after Hagedorn and hadronic decay cascades for a potential \HS with $m=10\GeV$. The overall normalization is arbitrary fixed to the experimental proton value. Experimental data by ALICE \cite{Abelev:2013vea,Abelev:2013xaa,Abelev:2013zaa,Adam:2015yta,Adam:2015vda,Acharya:2017bso}. Results are shown for two different values $T_H=167\MeV$ and $152\MeV$ for the \HT.
    }
    \label{fig:AlicePlotMulti}
  \end{center}
\end{figure}

It is now worth comparing these relative yields with experimental yields. We take here the high-energy LHC data measured by ALICE \cite{Abelev:2013vea,Abelev:2013xaa,Abelev:2013zaa,Adam:2015yta,Adam:2015vda,Acharya:2017bso}.  A comparison of the relative yields of the decays of a \HS with $m=10\GeV$ with the experimental data is shown in \cref{fig:AlicePlotMulti}.
Here, and also in all following figures, the absolute normalisation will be fixed to the experimental proton yield.  As can be seen, a single \HS with large mass is not able to describe the experimental multiplicities; the distribution is too hard. Even with a bootstrap with $R=1.2\fm$ and thus a (smaller) \HT of $T_H=152\MeV$, as described above, higher mass states, especially the light nuclei, are overestimated. Only a reduction of the \HT further down to even lower values could yield a satisfactory description. Anyhow, this can only be achieved by further increasing the \HS size \cite{Beitel:2014kza}.

While here the mass distribution looks thermal, it is only governed by the \HT. Thus a SOC prescription may lead to thermal (looking) yields. Concluding from \cref{fig:AlicePlotMulti}, the intrinsic \HT leads to a mass dependence, which is too hard. This statement relies on the results of the ad-hoc mass choice of $m=10\GeV$. Since the branching rations are only mass independent on a logarithmic scale, the yields could change by looking into them with some detail and some changes of the mass. Nevertheless, a qualitative change of the picture is not expected. Therefore, we conclude this section with the statement, that within our \HS decay scenario, a scale invariant decay of \HS results in particle yields which are too hard, \ie show a slope parameter, which is too large.

\section{Thermal Hagedorn gas}
\label{sec:thermalGas}

\begin{figure}[htb]
  \begin{center}
    \hspace*{\fill}%
    \includegraphics[width=0.95\columnwidth,clip=true]{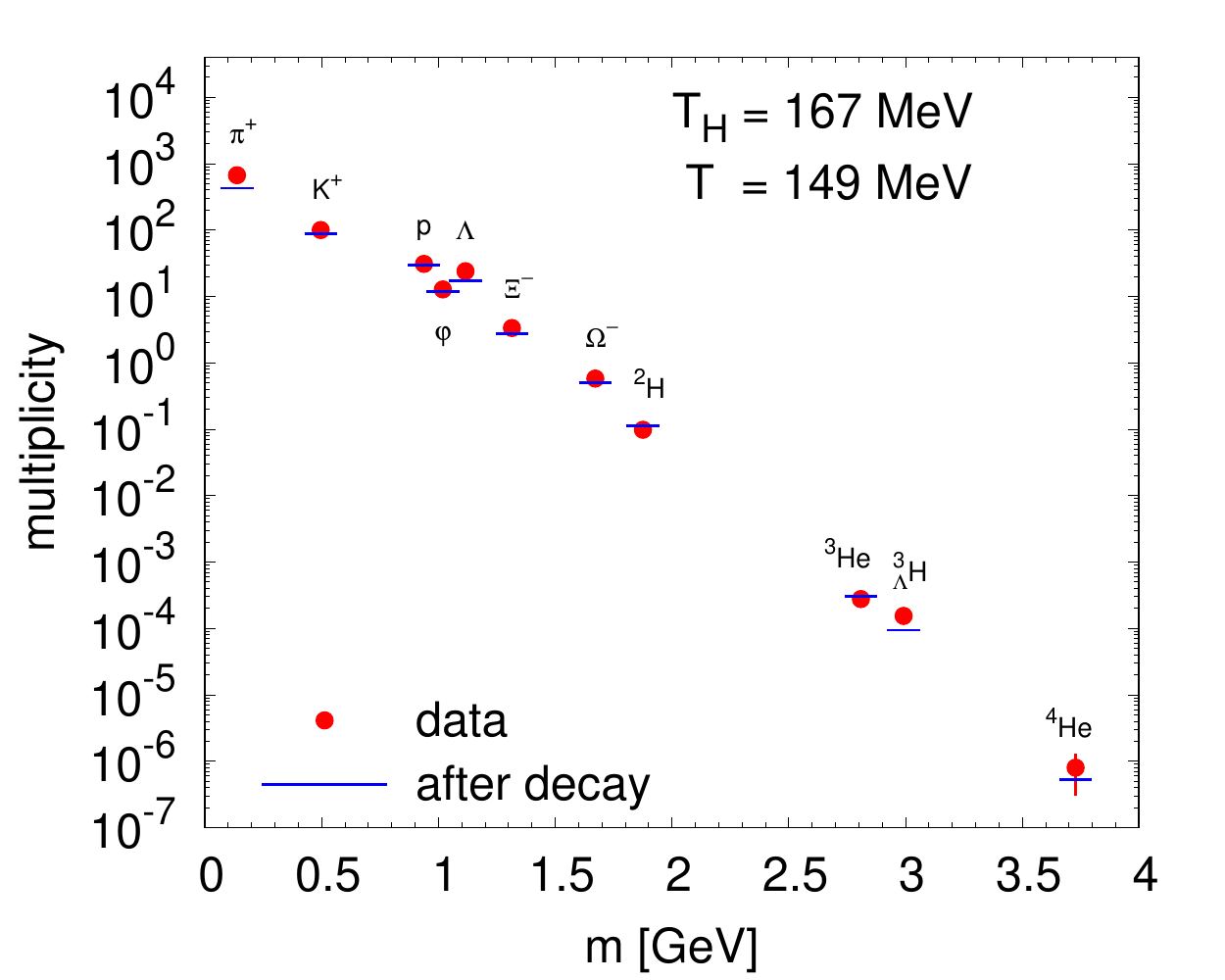}
    \hspace*{\fill}%

    \caption{
      Multiplicities as in \cref{fig:AlicePlotMulti}, but now for a gas of thermalized ($T=149\MeV$) hadron resonances and \HSS ($T_H=167\MeV$).
    }
    \label{fig:AliceThermal}
  \end{center}
\end{figure}

Turning to the picture of a thermalized gas of hadronic and Hagedorn resonances, an additional degree of freedom is introduced by the temperature of the system. An integration of the \HS mass spectra weighted by the Boltzmann factor for a given mass $m$,
\begin{align}
  n(m,T)=\frac{4\pi}{(2\pi)^3} m^2T\K2(m/T)\ ,
  \label{eq:BoltzmannFak}
\end{align}
with $\K{n}$ indicating modified Bessel functions, is necessary. Then the hadronic feed down of these thermal averaged \HSS has to be calculated. A fitting procedure applied after the decays to fit the experimental data of protons and light nuclei yields a temperature of $T=149\MeV$ for $T_H=167\MeV$ and $T=144\MeV$ for $T_H=152\MeV$.  The results for the first setup are shown in \cref{fig:AliceThermal}; the differences to the second setup are nearly invisible.

Some comments are in order. First, the number of mesons included in GiBUU is quite low compared to \eg UrQMD \cite{Bass:1998ca,Bleicher:1999xi}, or PDG \cite{Tanabashi:2018oca}. Therefore also the mesonic yields in these \HS decays may be underrepresented. This is why we only used the baryonic sector to fix the temperature. Second, the yields of the strange mesons and baryons are shown as is; no strangeness suppression factor has been applied. Third, these fits are meant to present the overall success. These fits are not intended to be high precision fits; therefore we just provide the resulting temperatures and abstain to give $\chi^2$ values.

It is obvious, that due to the additional degree of freedom the agreement of the model is much better than in the previous section. But also the production channels of the different particles is qualitatively different. In the picture of a thermalized gas, one has a thermal contribution of stable particles, while also the feed down from decays of higher lying resonances contributes.  So only approx.~half of the final $^{2}$H may be claimed to be (directly) thermal, while the other half stems from decays of \HSS. In the case of the higher masses of the nuclei, the situation is even more extreme: only approx.~20\proz{} of $^4$He are thermal, while 80\proz{} stem from feed down.
(Interestingly, this finding seems to depend on the underlying \HT; for the lower \HT $T_H=152\MeV$, the relative contribution of the feed-down decays is much larger.)

Therefore it would be worthwhile to study the influence of higher lying resonances of the nuclei, as \eg in ref.~\cite{Vovchenko:2020dmv}. There the importance of these higher resonances was limited to a level of 5\proz at high energetic collisions at LHC and had a sizable effect for low energetic collisions with large baryochemical potentials.
If one would expand the model presented here by all these higher resonances, one would also expect a large occupation of these states, which would then lead to sizable contribution to the yields of stable nuclei after hadronic feed down. Anyhow, as mentioned above, this may be left for a future study.

\begin{figure}[htb]
  \begin{center}
    \hspace*{\fill}%
    \includegraphics[width=0.95\columnwidth,clip=true]{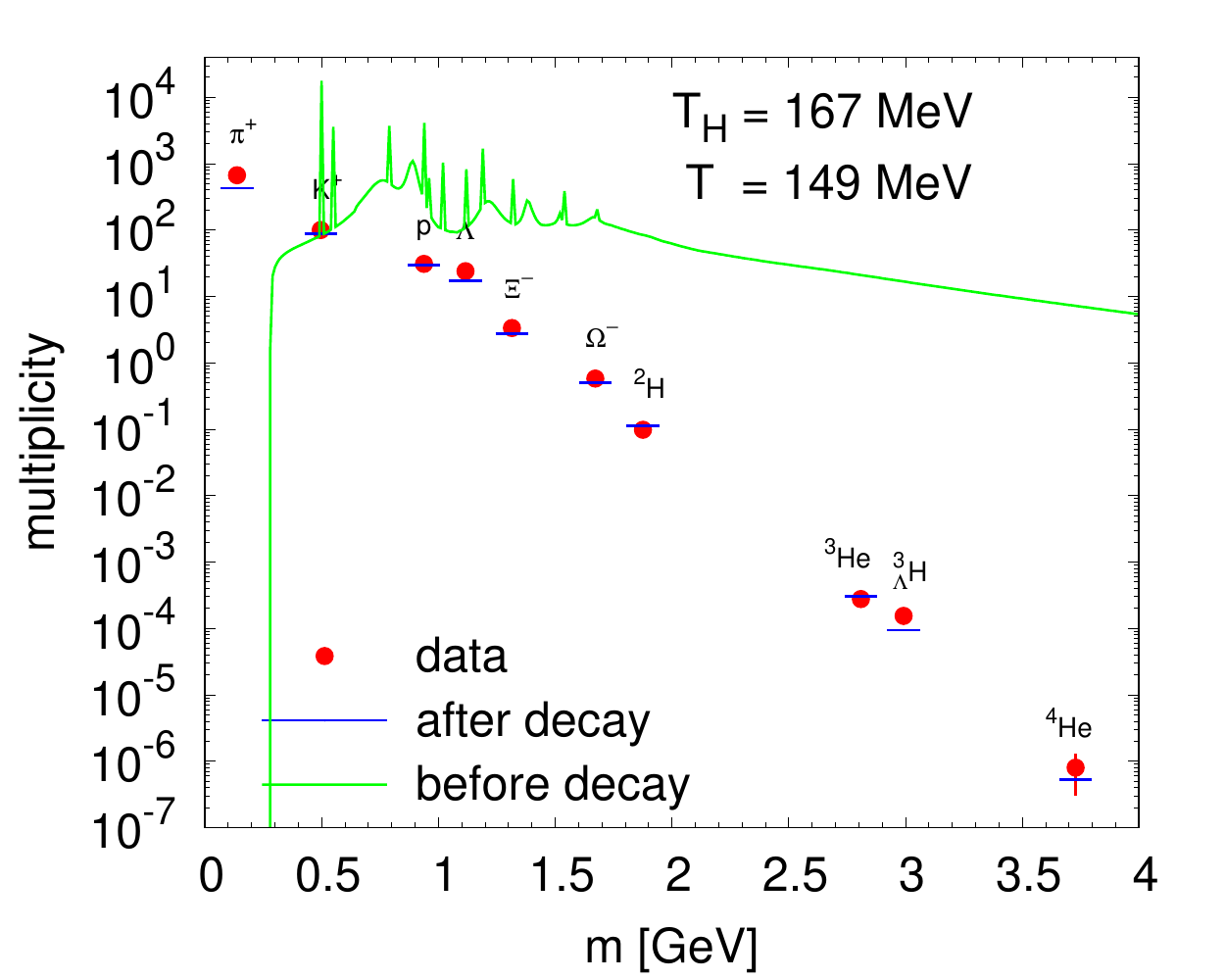}
    \hspace*{\fill}%

    \caption{
      As \cref{fig:AliceThermal}, but here is also shown the mass distribution of all potential \HSS before decays.
    }
    \label{fig:AliceThermal0}
  \end{center}
\end{figure}

In order to illustrate, that the final yield of stable particle is far from the spectrum of \HSS before decay, we indicate in \cref{fig:AliceThermal0} this spectrum in comparison to the final yields.
It is worth to emphasize, that here the normalization both of the spectra `before decays' and `after decays' are the same and the number of \HS with masses comparable to that of \eg $\alpha=^4$He are indeed seven orders of magnitude larger.

\section{Cooling of a Hagedorn gas with the constraint of chemical non-equilibrium}
\label{sec:cooling}

One may apply the same criticism to the Hagedorn gas picture as to a thermal model relying on a hadron resonance gas alone: how could these loosely bound states survive at these temperatures?

We thus will follow the arguments in \cite{Vovchenko:2019aoz}, where the Saha equation is the natural explanation how thermal yields behave under the cooling of the system. The assumption that during the cooling of the system the yields of stable particles are frozen at the `chemical freeze-out' (most important for nucleons and antinucleons), chemical potentials for all resonances are fixed in their temperature dependence.
While in \cite{Vovchenko:2019aoz} it was possible to calculate a full `decay matrix', this is more involved for the prescription presented here, since one would have to tabulate the decays of all quantum number states $(BSI)$ with mass $m$ into stable hadrons. This asks for the extension of \cref{eq:precalc} from light nuclei to all stable hadrons. While possible in principle, it is a challenging task due to computer memory constraints and not yet feasible.

\begin{figure}[htb]
  \begin{center}
    \hspace*{\fill}%
    \includegraphics[width=0.95\columnwidth,clip=true]{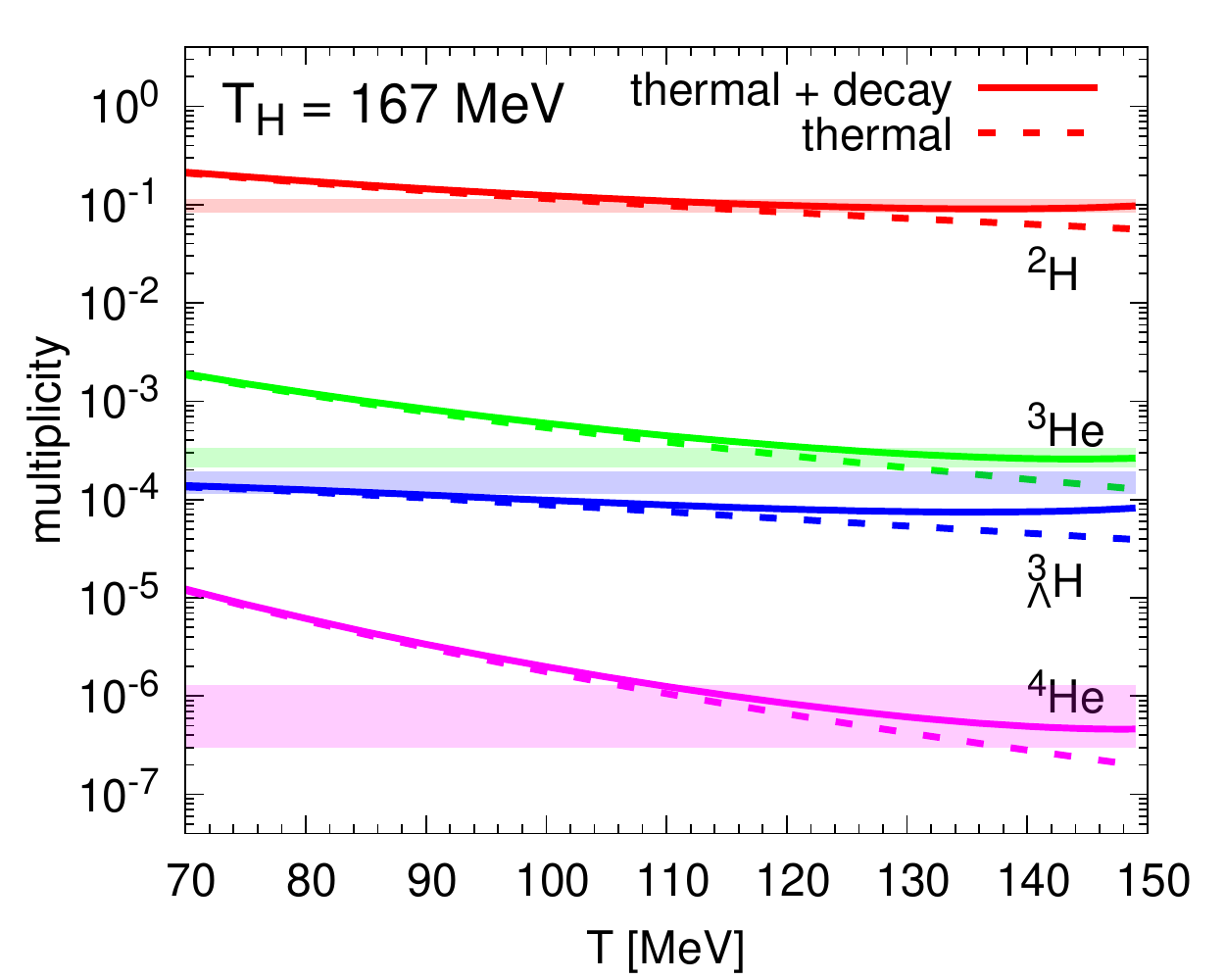}
    \hspace*{\fill}%

    \caption{
      The yields of light nuclei, when the yields of stable nucleons (protons and neutrons) and anti-nucleons are fixed to $T_{\rm cfo}=149\MeV$ with $T_H=167\MeV$. Solid lines indicate the total yields, while dashed lines show the contribution of thermal particles only. The colored bands indicate the experimental error bars of the data by the ALICE collaboration.
    }
    \label{fig:AsTemp}
  \end{center}
\end{figure}

Instead we apply a simplified setup, where we adjust the chemical potentials \textit{before} feed down and restrict to baryon number. We introduce a chemical potential for the absolute value of the particles baryon and anti-baryon number, $\mu_{|B|}(T)$, \ie both the number of protons and antiprotons are anchored, while the yields of the other stable particles are not considered. In this case, with particle numbers given by \cref{eq:BoltzmannFak}, the chemical potential is fixed by
$\exp(\mu_{|B|}(T)/T)=T_{\rm cfo}\K2(m_N/T_{\rm cfo})/(T\K2(m_N/T))$,
where $m_N=0.938\GeV$ stands for the nucleon mass and $T_{\rm cfo}$ indicates the chemical freeze-out temperature.
The resulting yields of light nuclei are displayed in \cref{fig:AsTemp}.
Within this picture, the yields of the light nuclei are (nearly) constant as function of the final temperature within some certain range. With decreasing temperature all yields start to increase. This behavior is not so pronounced for hyper-triton as for the other nuclei. Here the lack of introducing a chemical potential for the strange sector is visible.

This overall behavior has to be confronted with that of the Boltzmann factors \cref{eq:BoltzmannFak}, which would govern the temperature behavior otherwise and lead to a nearly exponential dependence of the yields as function of temperature (see the discussion in \cite{Vovchenko:2019aoz}).

Also shown in \cref{fig:AsTemp} is the relative contribution of thermal particles to the overall yield. With decreasing temperatures, the relative importance of feed-down particles vanishes.  This may be of interest, since it is well known since the results of \cite{Schnedermann:1993ws}, that the decay products of thermally distributed particles are not thermal, but look effectively cooler (the slope is steeper). Therefore a deeper inspection of the slopes of the decay products could lead to new insights about the production mechanism. Anyhow, this is beyond the possibilities of our approach, where only the absolute numbers of the light nuclei are accessible by the method relying on the tabulation according \cref{eq:precalc}.

\begin{figure}[htb]
  \begin{center}
    \hspace*{\fill}%
    \includegraphics[width=0.95\columnwidth,clip=true]{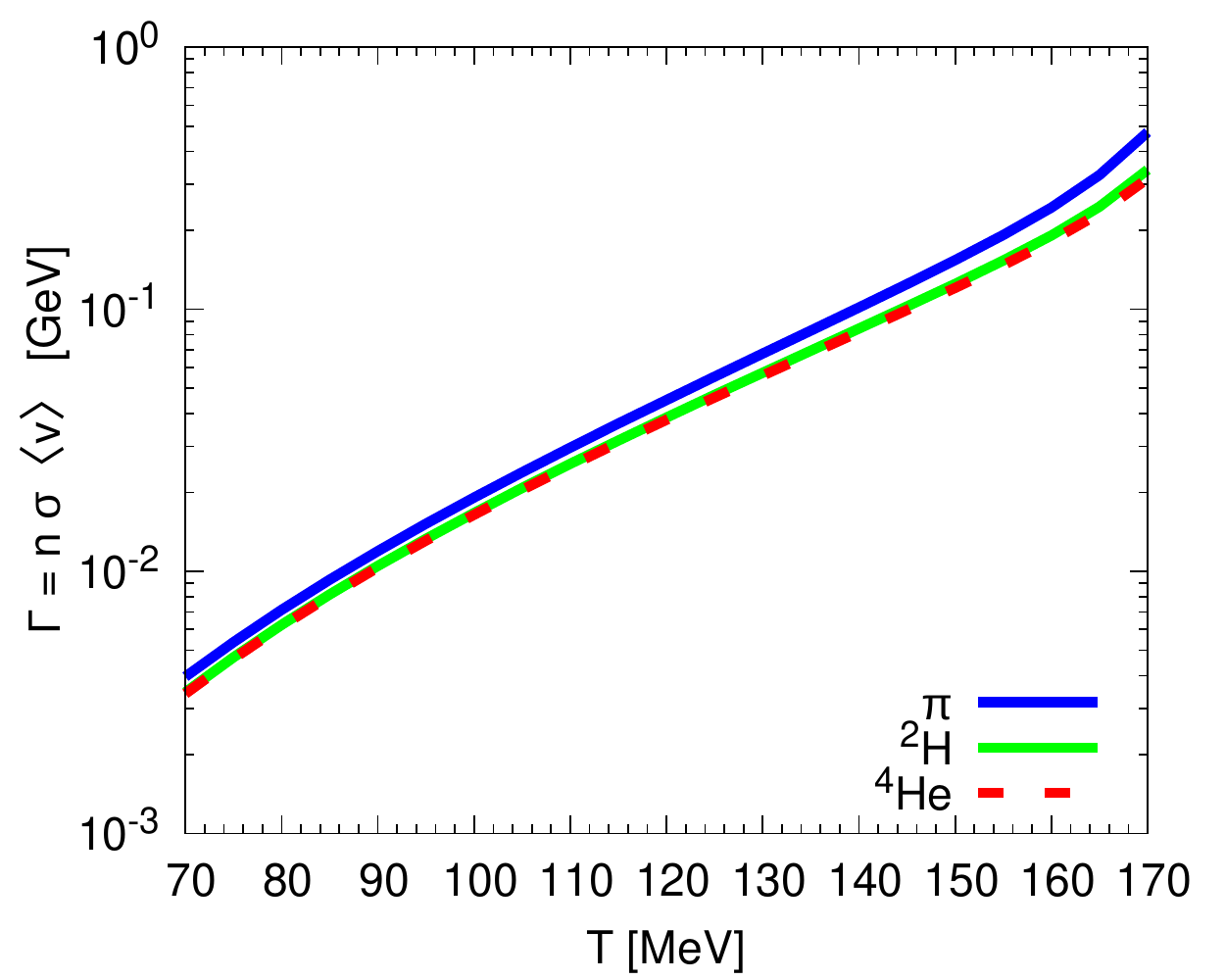}
    \hspace*{\fill}%

    \caption{
      The collision rate of pions, ${\rm d}=^2$H, and $\alpha=^4$He as function of temperature of the \HS gas ($\sigma=30\mb$).
    }
    \label{fig:Rate}
  \end{center}
\end{figure}

In order to justify the Saha equation picture also within the \HS prescription, we show in \cref{fig:Rate} the interaction rate of specific particles within a \HS gas.
Mass differences show up in slightly different curves.
It is now worth to realize some numbers.
A value for the rate of $\Gamma=0.2\GeV$, as realized for temperatures $T=150$-$160\MeV$, directly translates in lifetimes $\tau=1/\Gamma \simeq 1\fm$ and also represents the timescale of chemical equilibration of the \HSS.
The given interaction rates guarantee, that creation and destruction of the light nuclei proceed in relative chemical equilibrium after the chemical freeze-out.
Please note, that this interaction rate is governed mainly by the total \HS density. The introduction of $\mu_{|B|}(T)$ according the Saha picture only slightly changes the \textit{total} density.

On the other hand, the binding energies of the light nuclei are in the region $2.2$-$28.3\MeV$.
Only a quantum mechanical treatment of the creation and disintegration of the (tightly) bounded light nuclei in an open thermal system can lead to definite conclusions, when in the evolution of the fireball the light nuclei appear as bound states. This remains an outstanding question.
%

\section{Conclusions}
\label{sec:Conclusions}

Using directly a \HS prescription devoloped during the recent years can not allow to calculate decay rates into very rare channels as \eg into light nuclei. Relevant relative decay branchings may go down to $10^{-9}$, which is below any usual statistics available in Monte Carlo calculations by a factor ${\cal O}(10^5)$.  For the decays of \HSS with given quantum numbers and masses, a tabulation according to the usual bootstrap has been developed and allows to access these low yields. Since this tabulation only covers the number of particles, no other observable than the yields may be calculated in this way; quantities like energy spectra or flow still stay beyond reach.

The (relative) branching of \HSS into stable hadrons and light nuclei shows up to be nearly independent of the mass of the parent particle. Still, mass thresholds influence the yields and the above statement holds only true on a level, where the yields are depicted with a logarithmic scaling. The most general scaling behavior is reached for an averaging over all possible quantum numbers without any chemical potentials. Only this case is covered in this work.

The relative branchings are comparable with the experimental yields of the ALICE experiment. It shows that the \HS decays lead to an over-prediction of heavy mass states. Even lowering the \HT within reasonable ranges does not allow for a successful agreement.
Therefore the assumption of a scale-free system of \HSS is not sustained by our prescription, since the \HT is still too high compared to experimental data.

On the other hand, the introduction of an additional degree of freedom by assuming a thermalized system of \HSS, where in addition to the \HT also the temperature of the gas sets a scale, a satisfactory description of the experimental yields is achievable. With different values of the \HT, different temperatures yield the same level of accuracy of agreement.

As in a hadron resonance gas picture, a production of the light nuclei at chemical freeze-out temperature within the \HS gas suffers the same argument of having too small binding energies compared to the temperature. Taking the notion of `chemical freeze-out' seriously, all yields of stable particles are fixed at this point. Therefore a cooling below this temperature has to be considered akin to the Saha equation; chemical potentials of the stable hadrons influence those of the unstable once. In the present work, a simplified prescription of using a chemical potential for the absolute value of baryon and anti-baryon number has been shown.
Even in such a exploratory picture, the final yields of the light nuclei do only depend marginally on the final temperature, when staying within some range (as proposed in \cite{Vovchenko:2019aoz}).

A temperature dependence may be observed when looking at the ratio of `thermal' over `all particles'; if the final temperature is higher, the contribution of feed down particles may be larger. This could maybe be attacked by looking theoretically at the energy spectra of the particles. Anyhow, these spectra are beyond the given analysis. Also, only a description of experimental spectra using a realistic flow profile could really pin down that point.

In the present work, only high energetic heavy ion collisions have been covered. Here only the the thermodynamical properties of the \HS gas developed in our prescription are used.
Looking at the (very) low energy side, as \eg HADES at GSI, the full dynamical machinery implemented in the transport code may be used and there, also spectra of light nuclei may be calculated, maybe even with respect to the centrality of the collisions. This is left for future studies.

\begin{acknowledgement}

This work was supported by the Bundesministerium f\"ur Bildung und
Forschung (BMBF), grant No.~3313040033.

\end{acknowledgement}

\bibliographystyle{epj}
\bibliography{main}

\end{document}